\documentclass{aa}
\usepackage{graphics}

\begin{document}
\newcommand{\kms}{\,{\rm km}\,{\rm s}^{-1}}
\newcommand{\mum}{$\,\mu$m}
\newcommand{\muk}{$\,\mu$K}
\newcommand{\GHz}{$\,$GHz}
\newcommand{\mJy}{$\,$mJy}
\def\simgt{\mathrel{\raise0.35ex\hbox{$\scriptstyle >$}\kern-0.6em
\lower0.40ex\hbox{{$\scriptstyle \sim$}}}}
\def\simlt{\mathrel{\raise0.35ex\hbox{$\scriptstyle <$}\kern-0.6em
\lower0.40ex\hbox{{$\scriptstyle \sim$}}}}

\thesaurus{02(12.03.3; 12.03.4; 12.03.1; 13.09.1)}
\title{Implications of SCUBA observations for the Planck Surveyor}
\author{Douglas Scott\inst{1} \and Martin White\inst{2}}
\offprints{Douglas Scott}
\mail{Douglas Scott}
\institute{Department of Physics \& Astronomy, University of British
Columbia, Vancouver, B.C. V6T~1Z1, Canada (dscott@astro.ubc.ca)
\and
Departments of Astronomy and Physics, University of Illinois at
Urbana-Champaign, IL~61801, USA (white@physics.uiuc.edu)}
\date{Received date / Accepted date}
\titlerunning{SCUBA and Planck}
\authorrunning{D. Scott \& M. White}
\maketitle

\begin{abstract}
We investigate the implications for the Planck Surveyor
of the recent sub-millimetre number counts obtained using the SCUBA camera.
Since it observes at the same frequency as one of the higher frequency
science channels
on Planck, SCUBA can provide constraints on the point-source contribution
to the CMB angular power spectrum, which require no extrapolation in frequency.
We have calculated the two-point function of these sub-millimetre sources,
using a Poisson
model normalized to the observed counts.  While the current data are uncertain,
under reasonable assumptions the point-source contribution to the anisotropy
is comparable to the noise in the 353\GHz\ channel.
The clustering of these sources is currently unknown, however if they
cluster like the $z\sim 3$ Lyman-break galaxies their signal would be larger
than the primary anisotropy signal on scales smaller than about 10 arcminutes.
We expect the intensity of these sources to decrease for wavelengths longward
of $850\mu$m.  At the next lowest Planck frequency, 217\GHz, the contribution
from both the clustered and Poisson terms are dramatically reduced.  Hence
we do not expect these sources to seriously affect Planck's main science
goal, the determination of the primordial anisotropy power spectrum.  Rather,
the potential determination of the distribution of sub-mm sources is a
further piece of cosmology that Planck may be able to tackle.
\keywords{Cosmology: observations -- Cosmology: theory --
cosmic microwave background -- Infrared: galaxies}
\end{abstract}

\section{Introduction}

The study of the anisotropy in the Cosmic Microwave Background (CMB) has
the potential to teach us a great deal about the background cosmology in
which we live, about the formation of
structure and about the early universe.
Because of this promise, ESA selected the
Planck Surveyor\footnote{http://astro.estec.esa.nl/Planck/}
as the third Medium
sized mission of its Horizon 2000 Scientific Program.
With its wide range of frequencies, superb angular resolution, and high
sensitivity, Planck
has been hailed as the definitive CMB anisotropy experiment.

While much of the cosmological information is expected to come from the
angular power spectrum of {\it primary\/} anisotropies,
the sky contains more than
this simple imprint of the inhomogeneities at last scattering.
In particular the brightness fluctuations should be dominated on the smallest
scales by point sources.
One of the collateral science goals of Planck is to produce an all sky
catalogue of such sources over a wide range of frequencies.
Recent observations suggest that there may be many more bright
sub-millimetre sources
than previously expected, and it is the purpose of this paper to explore the
impact of these findings on the Planck mission.
While several authors (see below) have looked at the one-point statistics of
these sub-mm galaxies, little has been done on the two-point function
(or power spectrum) of these sources.
It is this statistic which is most familiar to the CMB community, and which
we concentrate on here.

The most exciting recent observations in the sub-mm waveband have come
from the new Submillimeter Common-User Bolometer Array
(SCUBA; Holland et al.~\cite{Holland}) on the James Clerk Maxwell Telescope.
A combination of the properties of the sky and the galaxies themselves
make the SCUBA 850\mum\ filter the optimal one for cosmology.
This waveband corresponds
closely with the 353\GHz\ channel of the High Frequency Instrument (HFI)
of Planck.  The central frequencies are almost identical, although the
Planck bandwidth will be considerably larger ($\Delta\nu/\nu=0.25$ rather
than the $\simeq0.1$ of the SCUBA 850\mum\ filter).
SCUBA has now been used to make several deep
integrations which have detected distant sources at 850\mum\
(Barger et al.~\cite{Barger}, Eales et al.~\cite{Eales},
Holland et al.~\cite{Holletal},
Hughes et al.~\cite{Hughes}, Smail et al.~\cite{SIB}), and this has
radically altered our expectations for the importance of dusty galaxies
at high redshift.

A summary of the source count observations is provided in Table~\ref{tab:obs}.
These new data thus provide us with direct measurements of the number
density of bright sources (albeit currently only over small patches of
the sky, and a limited range of flux)
at a frequency directly relevant to the Planck science channels.
We shall work primarily at 353\GHz, though we shall also extrapolate these
counts to nearby frequencies using models of the spectral energy distribution
of the sources.

There have been several estimates of how many sub-mm
sources Planck might
be able to detect (e.g.~Bersanelli et al.~\cite{RedBook}), as well as estimates
of other one-point statistics, often referred to as `confusion noise'
(Condon~\cite{Condon}, Blain et al.~\cite{BIS}).
As far as the fluctuations are concerned,
several studies have already been carried out on the implications of
point sources for the measurement of CMB anisotropies.  Much of this work
dealt more specifically with radio galaxy contributions at low frequency
(e.g. Tegmark \& Efstathiou~\cite{TegEfs}), or dealt with far-IR sources,
but concentrated on the uniform background contribution and the correlation
function
(e.g.~Franceschini et al.~\cite{Fraetal}, Bond et al.~\cite{BCH},
Wang~\cite{Wang}, Gawiser \& Smoot~\cite{GawSmo}, Blain et al.~\cite{BIS}).
Very similar studies have also been carried out for the X-ray
background (see Yamamoto \& Sugiyama~\cite{YamSug}, and references therein),
the optical background (see Vogeley~\cite{Vogeley}, and references therein),
and more exotic backgrounds (e.g.~Scott~\cite{DDM}).
The most relevant work are the recent papers by
Toffolatti et al.~(\cite{Toffetal}) and
Guiderdoni et al.~(\cite{GHBMG}) which contain many useful results, and
estimates of confusion noise, although they were written before the new
series of SCUBA measurements.
Our approach here also differs from theirs in that our predictions are based
on straightforward extrapolations from observable properties.
We deliberately avoid any modelling of the complex galaxy evolution
process.

To calculate the impact of these new data on Planck it will be necessary
to model the number density of sources as a function of flux, $N(>S)$.
We shall take care to ensure that our model does not overproduce, when
integrated, the far-infrared background (FIB) light detected by
(Puget et al.~\cite{Pugetal}, Fixsen et al.~\cite{Fixsen}).
At 353\GHz\ the background is approximately
$0.14\pm0.04\,{\rm MJy}\,{\rm sr}^{-1}$, although this is not a directly
measured quantity, so the real error bar at 353\GHz\ may be larger.
We construct our fiducial model so that the integrated light from the
sources contribute essentially all of this background.

\begin{table}
\begin{tabular}{cccccc}
$S_{\rm cut}$ & N  & 95\% CL  &  Area             & Density  \\
(mJy)   &    &      & (Sq. Deg.)   & $10^3$/(Sq. Deg.) \\ \hline \\
 2.0      &  5 &(1.8,10.9)&$1.6\times10^{-3}$&
 $3.2^{+1.7}_{-1.2}$ \\ \\
 3.0      &  2 &(0.3, 6.4)&$2.5\times10^{-3}$&
 $0.8^{+0.7}_{-0.4}$ \\ \\
 2.8      & 12 &(6.4,20.2)&$6.8\times10^{-3}$&
 $1.8^{+0.6}_{-0.5}$ \\ \\
 4.0      &  6 &(2.3,12.3)&$2.4\times10^{-3}$&
 $2.5^{+1.2}_{-0.9}$ \\ \\
 5.0      & 5 &(1.8,10.9)& $6.0\times10^{-3}$&
 $0.8^{+0.4}_{-0.3}$ \\ \\
 8.0      &  4 &(1.2, 9.4)&$4.0\times10^{-3}$&
 $1.0^{+0.6}_{-0.4}$ \\
\end{tabular}
\caption{SCUBA point source observations at 850\mum\ (353\GHz).
We list the flux level to which each field was searched (generally 3 times
the rms level), the number of sources found, the Bayesian
95\% confidence level on
the mean counts coming from Poisson statistics, the area covered, the number
density of sources (with $\pm1\sigma$ errors also from Poisson statistics).
These numbers were taken from Hughes et al.~(1998), Barger et al.~(1998),
Eales et al.~(1998), Smail et al.~(1997), Chapman et al.~(in
preparation), and Holland et al.~(1998),
respectively.}
\label{tab:obs}
\end{table}

\section{Counts and the Far Infrared Background}

\begin{figure}
\resizebox{\hsize}{!}{\includegraphics{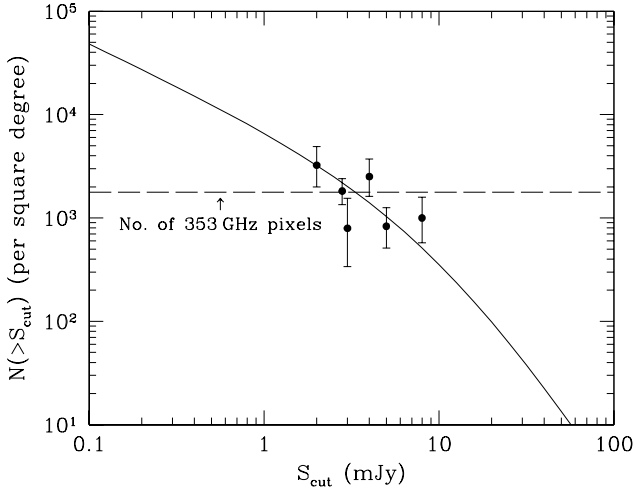}}
\caption{The number of objects brighter than flux $S_{\rm cut}$ as a
function of $S_{\rm cut}$.  The curve is
equation~(\protect\ref{eqn:Nfit}) with our fiducial parameters.
The data are taken from Table~\protect\ref{tab:obs}, with the second and third
points offset for clarity.  The error bars are $\pm1\sigma$ (68\% Bayesian
confidence region) errors based on
Poisson statistics.  The horizontal dashed line is an estimate of the
number of pixels on the whole sky for the Planck 353\GHz\ channel, assuming
10 pixels per $4.5'$ beam.}
\label{fig:counts}
\end{figure}

Guided by models of galaxy formation (e.g.~Toffolatti et al.~\cite{Toffetal},
Guiderdoni et al.~\cite{Guidetal}, Blain et al.~\cite{Blainetal}),
we model $N(>S_{\rm cut})$ as a double-power-law,
\begin{equation}
N(>S_{\rm cut}) = N_0 \left( {S\over S_0} \right)^{-\alpha}
  \left( 1 + {S\over S_0} \right)^{-\beta}
\label{eqn:Nfit}
\end{equation}
(see also Borys et al.~\cite{BorCS}),
where for convenience we take $S_0\,{=}\,10$\mJy.
Such a parameterization will certainly not be valid for all values of $S$,
but it will be adequate for our purposes.
Matching to the general behaviour of successful models, and
normalizing specifically to the Hubble Deep Field
(HDF) counts, we obtain `fiducial' values for these parameters of
$N_0=4.05\times 10^6\,{\rm sr}^{-1}$, $\alpha=0.8$ and $\beta=1.8$.
This fit, plus the data of Table~\ref{tab:obs}, are shown in
Fig.~\ref{fig:counts}.
Some models show the counts to steepen even more at the bright end, but
this has little effect on any results, since it is already steep enough
that any upper flux cut is not very important.
We also show in Fig.~\ref{fig:counts} an estimate for the number of
pixels over the whole sky at 353GHz, to indicate where we expect one
source per pixel.  We have assumed an oversampling by a factor of 10
pixels per beam as an illustrative number.  With our adopted source
counts model, Planck will then have about one 20\mJy\ source per beam
(which then sets the basic level of `confusion noise').

In practice Planck will only be able to detect individual sources with
at best a flux cut of $S\simgt100$\mJy\, using data from
the 353\GHz\ channel alone.
With extra information from the higher frequency channels (as well as
information from other instruments, at least in some regions of the sky),
it should be possible to remove all sources down to a few$\,\times\,10$\mJy.
The SCUBA counts constrain the model at somewhat lower flux levels than these,
however in our model the counts at the SCUBA flux levels contribute
significantly to the IR background and hence the CMB fluctuations if the
sources are clustered (see below).
In any case the precise flux cut for Planck is not currently easy to estimate,
and so we have erred on the side of conservatism; if the flux cut ends up
being higher than we are assuming here, then the fluctuations will only be
larger.

By describing the counts in this phenomenological way we avoid any direct
modelling of galaxy formation, evolution and spectral synthesis.  Currently
there are too many free parameters in these `semi-analytic' models to yield
a great deal of insight.  Instead we prefer to use simple model fits to
observables on the sky, which are motivated by the current data.  Because
of this we are considering only the two-dimensional distribution of objects
on the sky, with no requirement on the radial distribution.

\begin{figure}
\resizebox{\hsize}{!}{\includegraphics{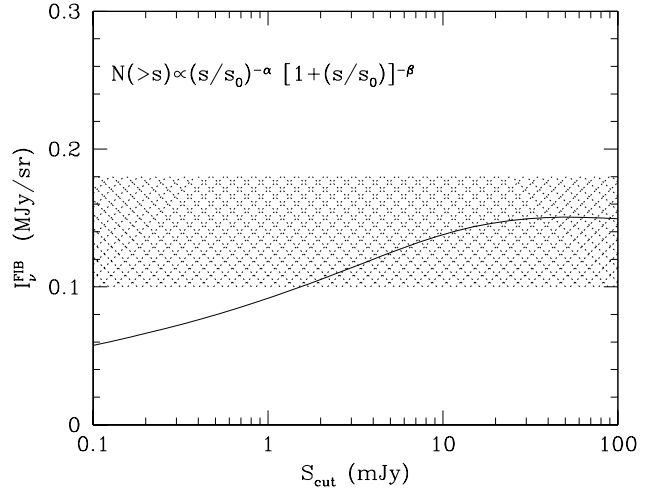}}
\caption{The integrated flux as a function of the upper flux cut,
with $S_0\,{=}\,10$\mJy, $\alpha\,{=}\,0.8$, $\beta\,{=}\,1.8$, and
normalized to the HDF counts.
The hatched region is the FIB detection of
Fixsen et al.~(\protect\cite{Fixsen}).  Note that in our fiducial model
much of the integrated flux comes from sources near the flux levels
detected by SCUBA.}
\label{fig:FIBscut}
\end{figure}

\begin{figure}
\resizebox{\hsize}{!}{\includegraphics{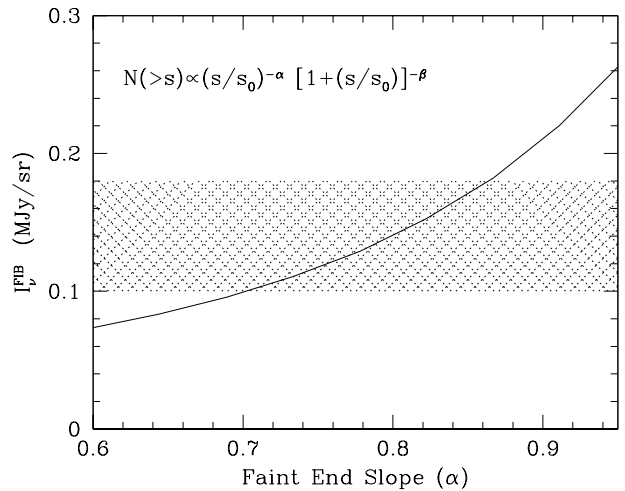}}
\caption{The total integrated flux as a function of the faint end slope,
$\alpha$ of equation~(\protect\ref{eqn:Nfit}), with the other
parameters fixed at their fiducial values.  The hatched region
is the FIB detection of Fixsen et al.~(\protect\cite{Fixsen}).}
\label{fig:FIBalpha}
\end{figure}

The contribution of these sources to the FIB is just the total flux per unit
solid angle, or
\begin{equation}
 I_{\nu}^{\rm FIB}
  = \int_0^{S_{\rm cut}} S_\nu\,{{\rm d}N\over {\rm d}S_\nu}\,dS_\nu,
\end{equation}
which can be integrated by parts to yield
\begin{equation}
  \int_0^{S_{\rm cut}}\!N\,dS_\nu
   - N(S_{\rm cut})\,S_{\rm cut}
\end{equation}
(a little care has to be taken with minus signs, since conventionally
$N\equiv N(S{>}S_{\rm cut})$).
The faint end limit for constant slope is just
$\alpha(1-\alpha)^{-1}N(>S_{\rm cut}) S_{\rm cut}$.
We show in Fig.~\ref{fig:FIBscut} the contribution to the integrated background
light as a function of $S_{\rm cut}$.
Notice that the sources at the flux levels probed by SCUBA contribute
significantly to the background.  As we will see below, it is those clustered
sources which contribute most to the background that may be of greatest
interest to us here.
In Fig.~\ref{fig:FIBalpha} we show the contribution to the FIB, integrating to
$S_{\rm cut}\,{=}\,\infty$, i.e.~the total background, as a function of
the faint-end slope $\alpha$.
The FIB was first detected by Puget et al.~(\cite{Pugetal}), and has recently
been measured by Fixsen et al.~(\cite{Fixsen}).  Their value is shown in
Figs.~\ref{fig:FIBscut}, \ref{fig:FIBalpha} as the hatched region.
In our fiducial model the sub-mm sources account for {\it all\/} of the FIB.

Let us acknowledge that the real situation may be more complicated.
The counts may come from a number of separate populations, and so of course
there could be features in the actual curve.  In addition there is the
possibility that some more diffuse emission contributes to the FIB, and
is not accounted for
in these counts.  Indeed there are some early indications (Hughes et al.~1998,
Borys et al.~1998) that the counts may be flatter at the faint end than the
form we have adopted.  Again we have been conservative here; lower faint-end
slopes would require higher overall normalization in order to match the
background, implying stronger fluctuations.

\section{Power spectrum for sources}

We shall quote our results in `temperature' units as is usual in CMB
anisotropy studies.  For fluctuations about a mean, the conversion factor
{}from temperature to intensity (or flux) is
\begin{eqnarray}
{\partial B_\nu\over\partial T} &=&
  {2k\over c^2} \left( {kT_{\rm CMB}\over h} \right)^2
  {x^4 {\rm e}^x\over ({\rm e}^x-1)^2} \nonumber\\
&  = & \left( {99.27\,{\rm Jy}\ {\rm sr}^{-1}\over \mu\,{\rm K}} \right)
  {x^4 {\rm e}^x\over ({\rm e}^x-1)^2},
\end{eqnarray}
where $B_\nu$ is the Planck function, $k$ is Boltzmann's constant,
$x\equiv h\nu/k_BT_{\rm CMB}=\nu/56.84\,$GHz is the
`dimensionless frequency' and
$1\,{\rm Jy}\,{=}\,10^{-26}\,{\rm W}\,{\rm m}^{-2}{\rm Hz}^{-1}$.

If we assume that the number of sources of a given flux is independent
of the number at a different flux, and if the angular two-point function
of the point-sources is $w(\theta)$, then the angular power spectrum,
$C_\ell$, contributed by these sources is
\begin{equation}
  C_\ell(\nu)=\int_0^{S_{\rm cut}}\ S_\nu^2\ 
  {{\rm d}N\over {\rm d}S_\nu} \,dS_\nu
  + w_\ell \left(I_{\nu}^{\rm FIB}\right)^2\!,
\label{eqn:cltot}
\end{equation}
assuming that all sources with $S{>}S_{\rm cut}$ are removed.
Here, $I_{\nu}^{\rm FIB}=\int S\, {\rm d}N/{\rm d}S\, dS$ is the background
contributed by sources below $S_{\rm cut}$ as before.
Following the conventional notation (essentially introduced by
Peebles~\cite{Peebles}), $C_\ell$ is
the Legendre transform of the correlation function $C(\theta)$ produced
by the sources and $w_\ell$ is the Legendre transform of $w(\theta)$:
\begin{eqnarray}
C(\theta) & = & {\displaystyle 1\over 4\pi} \sum_{\ell} (2\ell+1)C_\ell
  P_\ell(\cos\theta) \\
w(\theta) & = & {\displaystyle 1\over 4\pi} \sum_{\ell} (2\ell+1)w_\ell
  P_\ell(\cos\theta),
\end{eqnarray}
with $P_\ell(\cos\theta)$ the Legendre polynomial of order $\ell$.
The first term in equation~(\ref{eqn:cltot})
is the usual Poisson shot-noise term
(see Peebles~\cite{LSSU} \S46, or Tegmark \& Efstathiou~\cite{TegEfs}),
the second is due to clustering, assuming that the clustering is independent
of flux.

Integrating by parts the Poisson term can be rewritten
\begin{equation}
C_\ell(\nu)=2\int_0^{S_{\rm cut}}\!N\,S_\nu\,dS_\nu
 - N(S_{\rm cut})\,S_{\rm cut}^2.
\end{equation}
At low flux it becomes $\alpha(2-\alpha)^{-1}N({>}S_{\rm cut})S_{\rm cut}^2$,
if the counts have slope $\alpha$ at the faint end.

\begin{figure}
\resizebox{\hsize}{!}{\includegraphics{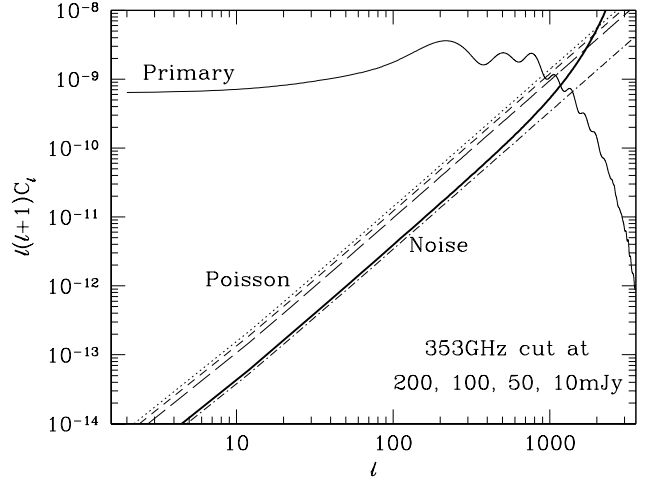}}
\caption{The Poisson component of the angular power spectrum of the point
sources, in dimensionless units, for a range of flux cuts
(i.e.~removing all sources brighter than $S_{\rm cut}$).  Higher flux cuts
give larger fluctuation levels.
To compare with the level of primary anisotropy expected, the prediction for
a standard CDM spectrum is also shown, normalized to COBE.  The thick solid
line is the expected contribution to the power spectrum from noise in the
353\GHz\ channel of the Planck HFI.}
\label{fig:cl}
\end{figure}

The shot-noise component of the angular power spectrum of our fiducial model
is shown in Fig.~\ref{fig:cl}, along with the primary CMB signal and the
projected noise in the Planck 353\GHz\ channel.
The Poisson fluctuations are calculated using our fiducial model and assuming
that we cut out all sources brighter than
$S_{\rm cut}\,{=}\,200, 100, 50, 10$\mJy.
Note that much of the fluctuation power comes from sources fainter than 50\mJy.
The uncertainty in the normalization of these curves is directly proportional
to our uncertainty in the counts, $N_0$, which we have normalized to the data
of Table~\ref{tab:obs}.
Clearly $N_0$ is uncertain to at least a factor of 2.  However, increasing
the normalization by this amount would overproduce the FIB unless the faint
end slope is also modified.

As an aside we mention that though the noise and Poisson component of the
sources appear to have similar power spectra, they are nonetheless quite
different entities.  The sources are on the sky, and thus contribute to
the flux in every observation of that pixel, whereas the noise varies
from observation to observation and by assumption is uncorrelated with the
signal in the pixel observed.  Given many observations of a given direction
on the sky (as expected for Planck),
the noise properties can be separated from the sky signal,
{\it even if\/} they have the same power spectra.  The estimated
instrumental noise can be subtracted from the measured power spectrum,
with any point source contribution being evident as an extra
$C_\ell={\rm constant}$ component.  We therefore expect that
analysis of the Planck data set will include fitting for an excess white noise
component, which would be most likely due to unclustered point sources.

\begin{figure}
\resizebox{\hsize}{!}{\includegraphics{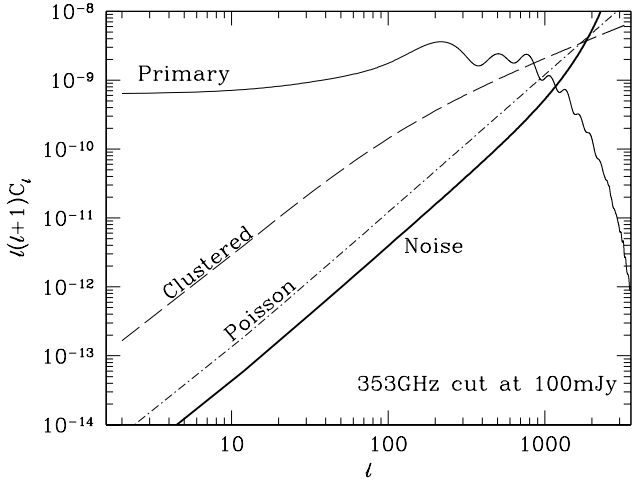}}
\caption{As in Fig.~\protect\ref{fig:cl}, showing a flux cut of 100\mJy\ only,
and including the component due to clustering.  We have modelled the
clustering using equation~(\protect\ref{eqn:wl}).}
\label{fig:cluster}
\end{figure}

We now turn to the other contribution to equation~(\ref{eqn:cltot}).
Unfortunately there is essentially no information about the clustering of
the SCUBA sources at present.  Hence the most conservative assumption would
be no clustering at all, but that is obviously unreasonable.  Hence,
although we tried to be as conservative as possible when discussing the
shot-noise power spectrum, for clustering we will just make some simple guess.
If we approximate $w(\theta)\propto\theta^{-\beta}$ then
$w_\ell\propto\ell^{\beta-2}$.  Assuming either that the sources cluster
like galaxies today or as Lyman-break galaxies
(Giavalisco et al.~\cite{GSADPK}) at $z\sim 3$,
we would expect $\beta\simeq0.8$--0.9.
As an illustrative example we can assume that the sources which make up
much of the FIB cluster like these Lyman-break galaxies at $z\sim 3$.
We suspect that this may in fact be close to reality, since it is
equivalent to assuming that the population is a highly biased one,
collapsing early.  On the other hand the SCUBA sources will span a wider
redshift range than galaxies selected by the UV-dropout technique, thereby
washing out the angular correlations to some extent.  But,
certainly the SCUBA-type sources are likely to
be more highly clustered than IRAS galaxies at low redshift.
Our example can be probably
considered an optimistic one in terms of clustered power.
If the objects are less biased than the Lyman-break galaxies (LBGs) by a factor
of $b^{-1}$ then one reduces $w_\ell$, and hence the clustered contribution
to $C_\ell$, by $b^{-2}$.

With our assumption
$w(\theta)\simeq \left(\theta/2^{\prime\prime}\right)^{-0.9}$ and
\begin{equation}
  w_\ell^{\rm LBG} \simeq 10^{-6} \left( {\ell\over 100} \right)^{-1.1}\!\!.
\label{eqn:wlLy}
\end{equation}
On large angles $w(\theta)$ is expected to drop below the power-law
behaviour assumed in equation~(\ref{eqn:wlLy}).  The scale of non-linearity
approximately marks this transition.  If we assume the power spectrum is
a power-law with index $n$ then $w_\ell\propto\ell^n$.  Thus we expect
that on larger angular scales $w_\ell$ will flatten and gradually turn over
to $w_\ell\propto\ell$.  To take this into account we cause $w_\ell$ to become
constant\footnote{Following our argument above there is no reason why $w_\ell$
might not drop to low-$\ell$, rather than go constant, but for $\ell\la 100$
the CMB signal dominates anyway, so our precise assumptions are not important.
If we cause $w_\ell$ to approach $\ell$ immediately then the clustered signal
is down from Fig.~\ref{fig:cluster} by an order of magnitude at $\ell=10$.
Thus we caution the reader that this simple model may overestimate the
clustered signal at low-$\ell$.  A more complete treatment would require
knowledge of the redshift distribution and the evolution of the clustering
of the sources.} for $\ell\la 100$, where we have chosen this multipole
because it is a round number, and not because we think it has any physical
significance.  Explicitly we approximate using
\begin{equation}
  {1\over w_\ell} = {1\over w_{100}^{\rm LBG}} + {1\over w_\ell^{\rm LBG}} \ .
\label{eqn:wl}
\end{equation}

We show the amplitude of the clustering signal in Fig.~\ref{fig:cluster}.
Note that the contribution due to source clustering dominates over the
Poisson term on the range of angular scales relevant to CMB anisotropies.
In fact if the clustering proves to be this strong then Planck may be able to
measure the power spectrum of the IR sources over a range of
angular scales.  We imagine that such a component will be included as one of
the foreground templates to be fitted for in a full Planck analysis -- the
template would include frequency dependance, as well as a power spectrum
which is white noise with perhaps one or two additional parameters to
describe the clustering part.  Exactly how to model this component
may change as we learn more from SCUBA and other instruments.
Further understanding of the clustering of these sources is clearly an
important direction for future research.

One other issue is the variance in these power spectra estimates.
Assuming that the point sources are a Poisson sample, it is straightforward
to estimate the cosmic variance associated with this component of the angular
power spectrum\footnote{If the full power spectrum is a sum of uncorrelated
terms, the variance of the total is just the sum of the variances.}.
The variance is a sum of two terms, one due to the finite number of modes
sampled by any given $C_\ell$ and the other from the Poisson nature of the
process.  So we have
\begin{eqnarray}
C_\ell &=& \int S_\nu^2\, {{\rm d}N\over {\rm d}S_\nu}\, dS_\nu, \\
{\rm and}\quad\left( \delta C_\ell \right)^2 &=&
   {2\over 2\ell+1} C_\ell^2 + \int S_\nu^4\,
   {{\rm d}N\over {\rm d}S_\nu}\, dS_\nu.
\label{eqn:dcl}
\end{eqnarray}
The first term of equation~(\ref{eqn:dcl}) is the usual result for Gaussian
fluctuations, the second term is the extra variance associated with the
Poisson sampling.
If only a fraction $f_{\rm sky}$ of the sky is observed then the first term
is increased by $f_{\rm sky}^{-1}$ while the Poisson term is unchanged.
For now it seems safe to assume that the uncertainty in our modelling of the
sources (i.e.~the normalization and shape of ${\rm d}N/{\rm d}S$)
is larger than the
estimate of equation~(\ref{eqn:dcl}), so we can neglect the latter.

\begin{figure}
\resizebox{\hsize}{!}{\includegraphics{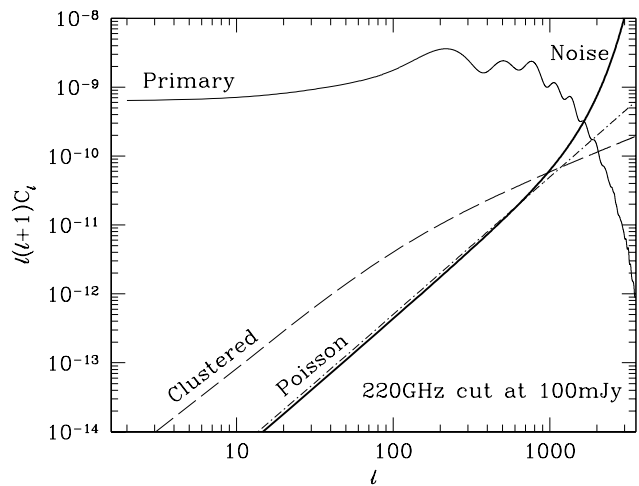}}
\caption{As in Fig.~\protect\ref{fig:cluster}, except at 217\GHz\ assuming
the fluxes scale as $\nu^{2.5}$.  We have kept the flux cut at 100\mJy\ to
isolate the effect of changing frequency, however in principle the higher
frequency channels could be used to isolate sources with 217\GHz\ flux
substantially less than 100\mJy.
This would lower the power spectra even further.}
\label{fig:cluster220}
\end{figure}

We expect the intensity of these sources to decrease for wavelengths
longward of $850\mu$m.
Assuming the flux decreases as $\nu^{2.5}$, similar to the slope of the
FIB, we have calculated the contribution at the next lowest Planck
frequency: 217\GHz.  As expected the contribution is considerably lower,
as shown in Fig.~\ref{fig:cluster220}. 
If we use information on the spectra of individual galaxies which have been
detected by SCUBA, (e.g.~Ivison et al.~1998,
Hughes et al.~1998),  then the derived slope may be steeper still, leading to
lower 217\GHz\ contributions.
Obviously at even lower frequencies, the signal will be correspondingly
reduced.
At the next higher frequency channel, 545\GHz, the sources are obviously
brighter.  We find that the Poisson contribution with a 100\mJy\ cut dominates
the noise by more than 2 orders of magnitude over the range $2\la\ell\la10^3$,
with the clustered contribution potentially larger still.

\section{Conclusions}

We have investigated the implications of the recent SCUBA number counts
for the Planck Surveyor.
Since it observes at the same frequency as one of the main science channels
on Planck, SCUBA can provide constraints on the point-source contribution
to the CMB angular power spectrum which require no extrapolation in frequency.
While previous authors have investigated mainly
one-point statistics of several
SCUBA fields, it is the two-point function which has the most impact on the CMB
science which will be done with Planck.
We have calculated the two-point function of point-sources, using a Poisson
model normalized to the observed counts.  While the current data are uncertain,
under reasonable assumptions the point-source contribution to the anisotropy
is comparable to the instrumental noise in the 353\GHz\ channel.  We have
emphasized that if the instrumental noise power spectrum can be accurately
estimated (as is expected to be the case for Planck, since each pixel is
observed multiple times), then even a white noise contribution from point
sources could still be detected.
The clustering of these sources is extremely uncertain, however if they
cluster like the $z\sim 3$ Lyman-break galaxies their signal would be larger
than the primary anisotropy signal on scales smaller than about 10 arcminutes.
We expect the intensity of these sources to decrease for wavelengths longward
of $850\mu$m.  At the next lowest frequency channel, 217\GHz, the contribution
from both the clustered and Poisson terms is dramatically reduced.

The bottom-line is that the sub-mm sources revealed by SCUBA will not have
a strong impact on the most important goal of the Planck mission, that of
precisely characterising the CMB anisotropy.  For the entire Low Frequency
Instrument, and the three lowest frequency channels of the HFI there will
be no significant contribution.  And certainly the signals in the higher
frequency channels can be used to remove point sources and recover most of
the CMB information even at 353\GHz.  Moreover, the possibility of actually
measuring the Poisson and clustering signals over most of the sky for these
galaxies provides Planck with yet another way of tackling fundamental
cosmological issues.

\begin{acknowledgements}
This work was supported by the Natural Sciences and Engineering Research
Council of Canada and by the National Science Foundation of the USA.
\end{acknowledgements}


\begin{thebibliography}{9}

\bibitem[1998]{Barger}
Barger A.J., Cowie L.L., Sanders D.B., Taniguchi Y.
 1998, Nature, 394, 248 [astro-ph/9806317]

\bibitem[1996]{RedBook}
Bersanelli M., et al.~1996, COBRAS/SAMBA, SCI(96)3, ESA, Paris

\bibitem[1998]{BIS}
Blain A.W., Ivison R.J., Smail I. 1998, MNRAS, 296, L29
[astro-ph/9710003]

\bibitem[1998]{Blainetal}
Blain A.W., Smail I., Ivison R.J., Kneib J.-P. 1998, MNRAS, in press
[astro-ph/9806062]

\bibitem[1991]{BCH}
Bond J.R., Carr B.J., Hogan C.J. 1991, ApJ, 367, 420

\bibitem[1998]{BorCS}
Borys C., Chapman S.C., Scott D. 1998, MNRAS, submitted [astro-ph/9808031]

\bibitem[1974]{Condon}
Condon J.J. 1974, ApJ, 188, 279

\bibitem[1998]{Eales}
Eales S., Lilly S., Gear W., Dunne L., Bond J.R., Hammer F., Le Fevre O.,
 Crampton D. 1998, ApJ, submitted [astro-ph/9808040]

\bibitem[1998]{Fixsen}
Fixsen D.J., Dwek E., Mather J.C., Bennett C.L., Shafer R.A. 1998,
 ApJ, 508, 123 [astro-ph/9803021]

\bibitem[1991]{Fraetal}
Franceschini A., Toffolatti L., Mazzei P., Danese L., De Zotti G.
 1991, A\&AS, 89, 285

\bibitem[1997]{GawSmo}
Gawiser E., Smoot G.F. 1997, ApJ, 480, L1 [astro-ph/9603121]

\bibitem[1998]{GSADPK}
Giavalisco M., Steidel, C.C., Adelberger, K.L., Dickinson, M.E., Pettini, M.,
 Kellogg, M. 1998, ApJ, 503, 543 [astro-ph/9802318]

\bibitem[1996]{GHBMG}
Guiderdoni B., Hivon E., Bouchet F.R., Maffei B., Gispert R. 1996,
 ``Unveiling the Cosmic Infrared Background'', ed. E. Dwek,
 IAP conference proceedings, 348, p.$\,202$

\bibitem[1998]{Guidetal}
Guiderdoni B., Hivon E., Bouchet F.R., Maffei B. 1998,
 MNRAS, in press [astro-ph/9710340]

\bibitem[1998]{Holletal}
Holland W.S., et al. 1998, Nature, 392, 788

\bibitem[1999]{Holland}
Holland W.S., et al. 1999, MNRAS, 303, 659 [astro-ph/9809122]

\bibitem[1998]{Hughes}
Hughes D.H., et al. 1998, Nature, 394, 241

\bibitem[1998]{Ivisonetal}
Ivison R.J., Smail I., Le Borgne J.-F., Blain A.W., Kneib J.-P.,
Bezecourt J., Kerr T.H., Davies J.K. 1998, MNRAS, 298, 583
[astro-ph/9712161]

\bibitem[1973]{Peebles}
Peebles P.J.E. 1973, ApJ, 185, 413

\bibitem[1980]{LSSU}
Peebles P.J.E. 1980, The Large-Scale Structure of the University,
Princeton University Press, Princeton

\bibitem[1996]{Pugetal}
Puget J.-L., et al. 1996, A\&A, 308, L5

\bibitem[1993]{DDM}
Scott D. 1993, MNRAS, 263, 903

\bibitem[1997]{SIB}
Smail I., Ivison R.J., Blain A.W. 1997, ApJ, 490, L5

\bibitem[1996]{TegEfs}
Tegmark M., Efstathiou G. 1996, MNRAS, 281, 1297

\bibitem[1998]{Toffetal}
Toffolatti L., Arg{\"u}eso G{\'o}mez F., De Zotti G., Mazzei P.,
 Franceschini A., Danese L., Burigana C. 1998, MNRAS,
 297, 117 [astro-ph/9711085]

\bibitem[1998]{Vogeley}
Vogeley M.S. 1998, ApJ, in press [astro-ph/9711209]

\bibitem[1991]{Wang}
Wang B. 1991, ApJ, 374, 465

\bibitem[1998]{YamSug}
Yamamoto K., Sugiyama N. 1998, PRD, submitted [astro-ph/9807225]

\end{thebibliography}
\end{document}